\documentclass[12pt]{article}   
\usepackage{helvet} 

\usepackage{graphicx}  
\usepackage{amssymb}
\usepackage{amsmath}
\usepackage{subcaption}
\usepackage[margin=1.0in]{geometry}

\usepackage{color}
                
\usepackage{hyperref}
\hypersetup{
    colorlinks=true,
    linkcolor=blue,
    citecolor=blue,
    filecolor= magenta,      
    urlcolor= magenta 
}

\newcommand{\J}{{\mathbf J}}

\newcommand{\Reyuls}{{I{\relax\kern-.3em}R}}

\newcommand{\es}{\epsilon_s}

\newcommand{\ez}{\epsilon_0}

\title{An Extension of  Goldman-Hodgkin-Katz Equations \\ by Charges from Ionic Solution and Ion Channel Protein}
 
\author{Dexuan Xie\thanks{Department of Mathematical Sciences,
University of Wisconsin-Milwaukee, Milwaukee, WI 53201, USA. E-mail address: dxie@uwm.edu. }
 }

\date{}

\begin{document}
\maketitle

\begin{abstract}
The Goldman-Hodgkin-Katz (GHK) equations have been widely applied to ion channel studies, simulations, and model developments. However, they are constructed under a constant electric field, causing them to have a low degree of approximation in the prediction of ionic fluxes, electric currents, and membrane potentials. In this paper, the equations are extended from the constant electric field to the nonlinear electric field induced by charges from an ionic solution and an ion channel protein. The extended GHK equations are also shown to  include the classic GHK equations as special cases. Furthermore, a novel numerical quadrature scheme is developed to estimate one major parameter, called the extension parameter, of the extended GHK equations in terms of a set of electrostatic potential values. To this end, the extended GHK equations become a bridge between the ``macroscopic" ion channel kinetics (i.e. ionic fluxes, electric currents, and membrane potentials) and  the ``microscopic" electrostatic  potential values across a cell membrane.  Developing methodologies and computational tools for generating a set of required electrostatic potential values becomes one important research topic in the application of  the extended GHK equations. One natural way to do so is to use a numerical solution of the one-dimensional Poisson-Nernst-Planck ion channel model that has been used to construct the extended GHK equations. In this paper, a nonlinear finite element iterative scheme for solving this model is developed and  implemented as a Python software package. This package is then used to do numerical studies on the extended GHK equations, the numerical quadrature scheme, and the nonlinear iterative scheme. Numerical results confirm the importance of considering charge effects in the calculation of  ionic fluxes. They also validate the high numerical accuracy of the numerical quadrature scheme, the fast convergence rate of the nonlinear iterative scheme, and the high performance of  the software package.
\end{abstract}


\section{Introduction}

The Goldman-Hodgkin-Katz (GHK) equations include three equations, called the GHK flux, electric current, and voltage equations, for the calculation of  ionic fluxes, electrical currents, and membrane potentials across a cell's membrane, respectively \cite{fall2002computational,HilleBook2001,keener1998mathematical}. They were derived by David E. Goldman and Nobel laureates Alan Lloyd Hodgkin and Bernard Katz in the 1940s \cite{goldman1943potential, hodgkin1949effect}. Since then, many studies have been done on the equations, along with  a number of modified GHK equations, making them to become one of  the most successful achievements in electrophysiology \cite{clay2009determining,didier2018membrane,frankenhaeuser1960sodium,garcia2000identification,iyer2004computational,jafri1998cardiac,tamagawa2019mathematical,ten2006alternans}.  As one basic tool for cell membrane study, the GHK equations have been included in most textbooks on electrophysiology (e.g. \cite{fall2002computational,HilleBook2001,keener1998mathematical}).

Charges from an ionic solution and an ion channel protein can induce a strongly nonlinear electric filed and in turn significantly affect ionic fluxes, electrical currents, and membrane potentials. However,  the GHK equations hold only for a constant electric filed since they do not consider any charge effect within an ion channel pore. Thus, they may have a low degree of approximation accuracy in the prediction of ionic fluxes, electrical currents, and voltages across a cell membrane. Even so, they are still widely applied to various ion channel studies, simulations, and model developments because they establish the connections of  ionic fluxes, currents, and voltages with intracellular and extracellular concentrations and a membrane potential in algebraic expressions. 

For example, in a mitochondrial dynamical system reported in \cite{beard2005biophysical,dash2008analysis}, the GHK flux equation is used to connect the concentrations from the inside of an inner  mitochondrial membrane (IMM) with those from the outside of the IMM. To this end,  intracellular and extracellular concentration can be treated as the variable functions of the dynamical system. Clearly, this dynamical system can be improved if the GHK flux equation is replaced by a modified GHK flux equation workable for a nonlinear electric field. Unfortunately, no such modified GHK equations existed currently. 

\begin{figure*}[t]
 \centering
       \begin{subfigure}[b]{0.3\textwidth}
                \centering
                \includegraphics[width=\textwidth]{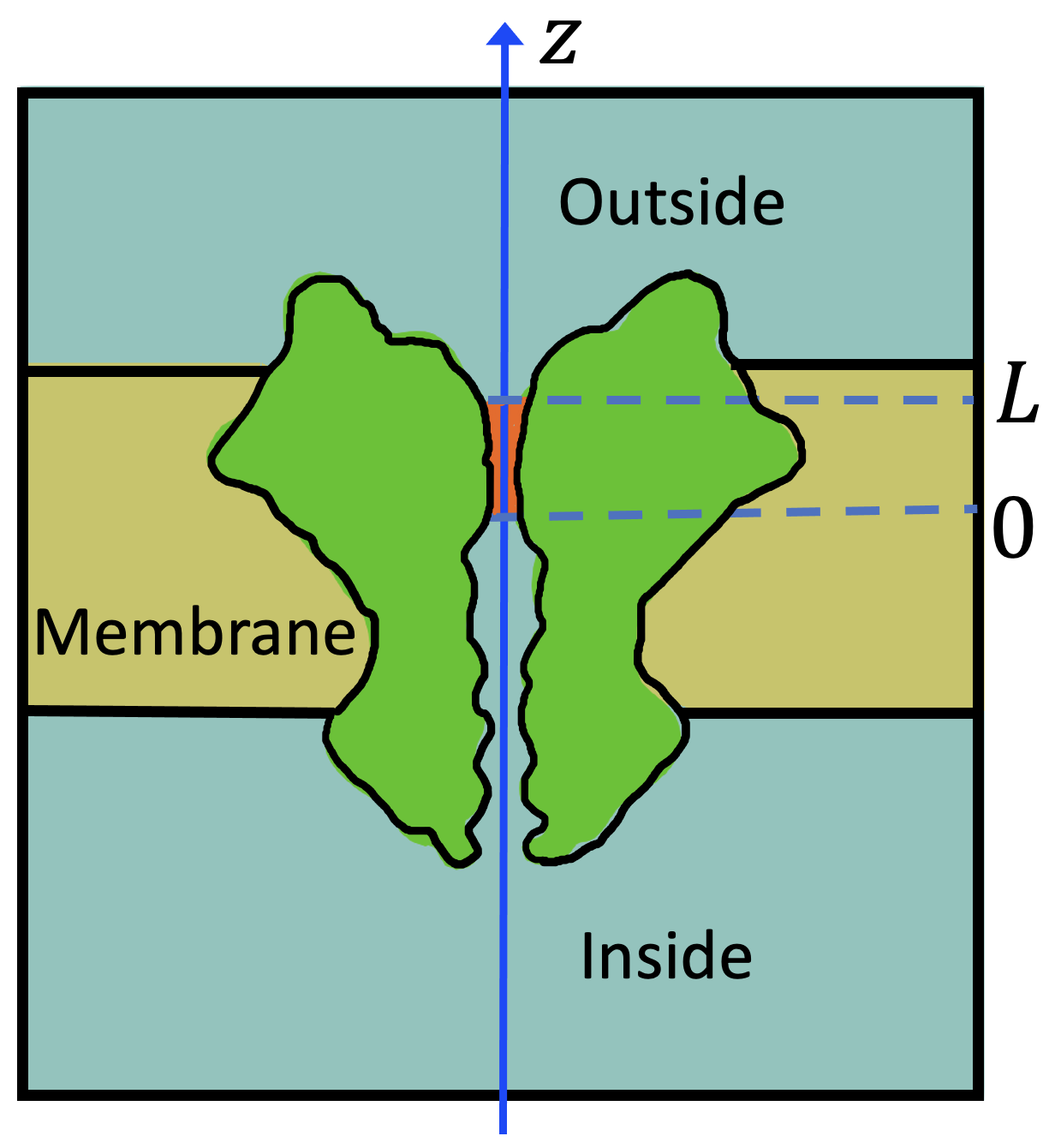}
                \caption{Selection 1}
        \end{subfigure}    
        \qquad \qquad
        \begin{subfigure}[b]{0.3\textwidth}
                \centering
                \includegraphics[width=\textwidth]{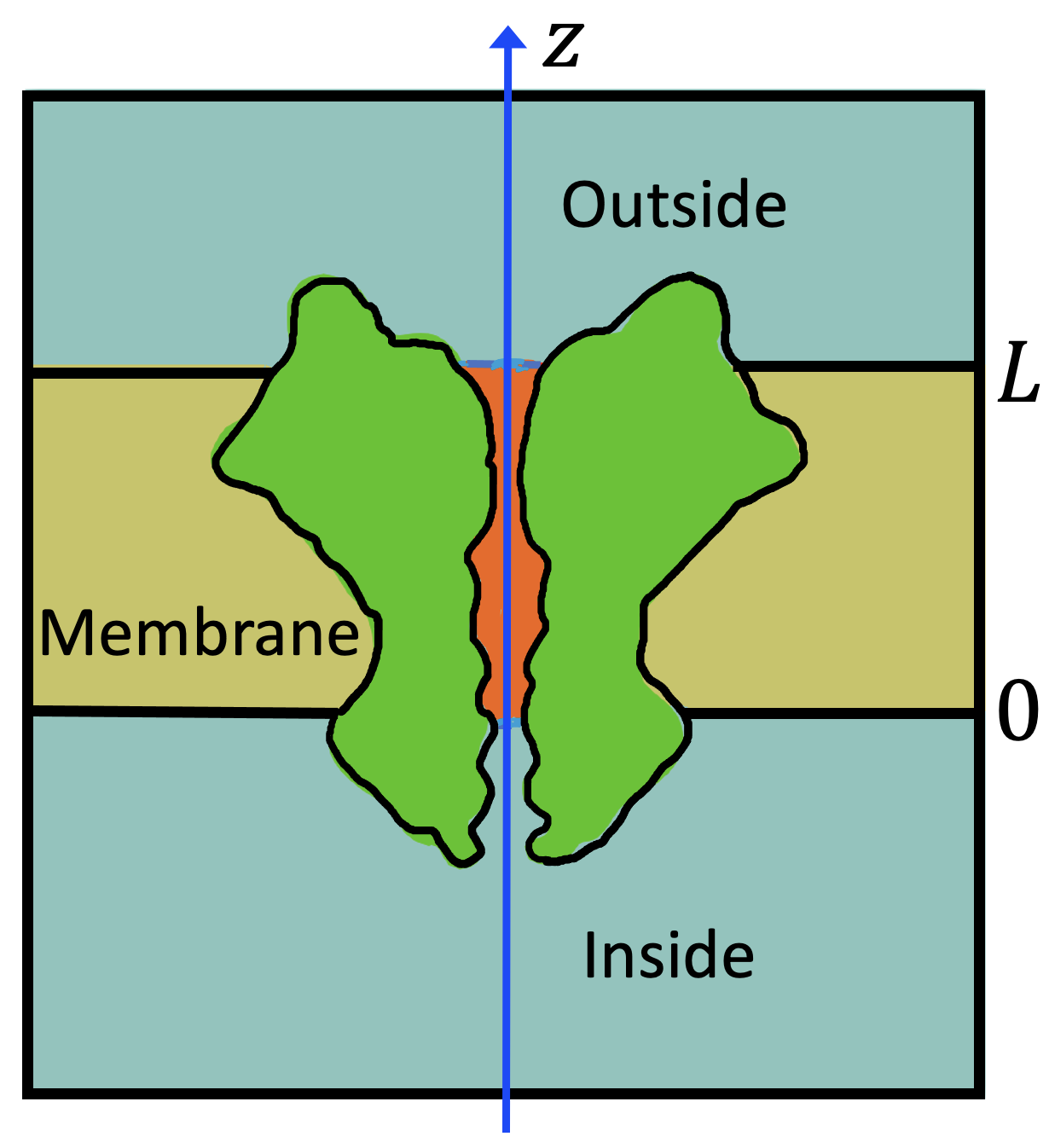}
                \caption{Selection 2}
        \end{subfigure}   
                \caption{Two typical selections of the ion channel interval $[0, L]$ for  the 1D PNPic model \eqref{1DPNPionChannel}.}
                                \label{channel_pore_intervals}
\end{figure*} 

On the other hand, lots of work was done on Poisson-Nernst-Planck ion channel (PNPic) models (e.g. see  \cite{LiXF2017,gardner2004electrodiffusion,liu2009one,nonner1998ion} for one-dimensional (1D) PNPic models and \cite{XiePNPic2021,Xie4PNPicPeriodic2020,Xie4PNPicNeumann2020} for recent  3D PNPic models). Although  they have a higher degree of accuracy than the GHK equations in the numerical calculation of ionic fluxes, electric currents, and voltages, these PNPic models cannot be used to substitute any GHK equation since they do not provide any algebraic formula similar to a GHK equation. Hence, extending GHK equations from a constant electric field to  a nonlinear electric field remains an open research problem. We intend to solve it in this work.  

To do so, we revisit the derivation of GHK equations based on a  typical 1D PNPic model (see \eqref{1DPNPionChannel} for its definition). We observe that  one key step in the derivation of the GHK flux equation is to treat  the electric field as a known coefficient function. In this way, a Nernst-Planck boundary value problem can be generated from the PNPic model and then solved analytically to obtain an algebraic formula for estimating ionic flux in terms of ionic concentration boundary values. However, expressing its solution, an ionic concentration function, in an analytical expression is difficult unless the electric field is set as a constant, as is done in the derivation of the GHK flux equation. In this work, we overcome the difficulty through introducing a new parameter, called the extension parameter, to account for the charge effects from an ionic solution and an ion channel protein. To this end, extending the GHK equations to a nonlinear electric field becomes possible.

Remarkably, we find that the extension parameter can be expressed as a definite integral of a composite function, $e^{Z_iu}$, over an ion channel pore interval (see \eqref{alpha-def} for details). Here, $Z_i$ denotes the charge number of ionic species $i$ and $u$ is a dimensionless potential function of the electric field. For a constant electric field, $u$ is simply a linear function. Thus, the extension parameter can be determined exactly. We can then reduce our extended GHK equations to the classic GHK equations. In this sense, the classic GHK equations can be regarded as special cases of our extended GHK equations. However, in the general case, we only can calculate the  extension parameter approximately using a numerical quadrature. While there exist several numerical quadratures including the composite trapezoidal rule that can work for such a calculation \cite{NA_Burden_Faires}, in this work, we will develop a new numerical scheme according to  the composite property of the integrand function $e^{Z_iu}$. That is,  we approximate $u$, instead of the integrand function $e^{Z_i u}$, as a piecewise linear interpolation function. In this way, the integral expression of  the extension parameter can be rewritten as a sum of the sub-integrals that can be evaluated exactly. Clearly, our new numerical scheme can have a higher degree of numerical  accuracy than the corresponding composite trapezoidal rule since its truncation error comes from a piecewise linear approximation of $u$ only. In contrast, the truncation error of the composite trapezoidal rule comes from the integrand function $e^{Z_i u}$.  

Since a set of electrostatic potential function values is required by our new scheme, we need a  methodology for its generation. One natural way for us to do so is to use a numerical solution of a 1D PNPic model. By far, several different 1D PNPic modes were developed and solved numerically (see \cite{LiXF2017,gardner2004electrodiffusion,liu2009one,nonner1998ion} for examples), which can be used to generate  the required electrostatic potential function values. However, we could not find their software packages from the public domain. Thus, we have to develop our own software package in order to carry out an initial study on our extended GHK equations. This turns out to be one important part of this work. To do so, we will develop a nonlinear finite element iterative scheme for solving the 1D PNPic model that we have used in the construction of our extended GHK equations based on our recent 3D PNPic work \cite{XiePNPic2021,Xie4PNPicPeriodic2020,Xie4PNPicNeumann2020}. We then will implement it in Python as a software package based on the  state-of-the-art finite element library from the FEniCS project \cite{fenics-book}. With our package, we will do numerical tests to confirm the  high numerical accuracy of our specially designed numerical quadrature, to demonstrate a fast convergence rate of our finite element iterative scheme,  and to show the importance of considering charge effects in the calculation of ionic fluxes. 

The remaining part of this paper is organized as follows: In Section~2, we revisit the GHK equations. In Section~3,  we present our extended GHK equations. In Section~4, we present our new numerical quadrature.  In Section~5,  we present our 1DPNPic finite element solver. In Section~6, we report numerical results.  Finally, conclusions are made in Section~7.

\section{Revisit of Goldman-Hodgkin-Katz equations} 
We set a normal direction of the membrane in  the $z$-axis direction of a 3D coordinate system, following what we did in our previous work \cite{XiePNPic2021,Xie4PNPicPeriodic2020,Xie4PNPicNeumann2020}. We then define  a dimensionless electrostatic potential density function, $u$, and an ionic concentration function, $c_i$, of species $i$ as  functions of one variable, $z$, on an interval $0\leq z \leq L$  by a 1D PNPic model as follows:

\begin{subequations}
\label{1DPNPionChannel}
\begin{eqnarray}
\label{electricField-Def}
- \es \frac{d^2 u}{dz^2}  = \frac{e_{c}^2}{\ez k_{B}T}  \sum_{i=1}^{n}Z_{i} c_i(z) + \rho(z), & 0<z<L,\\
\label{NP1D0}
\frac{d}{dz} {\cal D}_i \left[Z_i c_i(z) \frac{du(z)}{dz} + \frac{dc_i(z)}{dz}  \right] = 0,&0<z<L, \\
\label{boundary_values}
u(0)=u_0, \quad  u(L)=u_L, \quad
c_i(0) =c_{i,0},  \quad c_i(L) =c_{i,L}, &\qquad i=1,2,\ldots,n,
\end{eqnarray}
\end{subequations}
where $\rho$ denotes a permanent charge function;  ${\cal D}_i$ is a diffusion constant of species $i$; $n$ is the number of species in an ionic solution;  $\es$ denotes the water permittivity constant; $\ez$ is the permittivity of the vacuum;  $e_c$ is the elementary charge; $k_B$ is the Boltzmann constant; $T$ is the absolute temperature;  $Z_i$ denotes the  charge number of ionic species $i$; and $u_0$, $u_L$, $c_{i,0}$, and $c_{i,L}$ are the boundary values of $u$ and $c_i$, respectively. Here, \eqref{electricField-Def} is called a Poisson equation, \eqref{NP1D0}  a steady Nernst-Planck equation,  and \eqref{boundary_values} the Dirichlet  boundary value conditions. 

Two typical selections of the interval $[0, L]$ are illustrated in Figure~\ref{channel_pore_intervals}. Here, $L$ is the length of a portion of the ion channel pore, the end number $0$ is set inside the cell, and $L$ outside the cell.  In Selection~1, the interval $[0, L]$ corresponds to an ion-selectivity filter of an ion channel protein. In this case, the charge function $\rho$ can be positive for an anion selectivity filter or negative for a cation selectivity filter.

The dimensionless potential $u$ can be changed to an electrostatic potential, $\Phi$, in volts by
\begin{equation}
\label{u_Phi}
   \Phi(z) =  \frac{k_B T}{e_c} u(z), \quad 0<z<L.
\end{equation}
Here the constant $ \frac{k_BT}{e_c}$  is about  0.026 volts at $T=298.15$ Kelvin.

Clearly,  the steady Nernst-Planck equation \eqref{NP1D0} can be reformulated as
\begin{equation}
\label{NP1D2}
   \frac{d  \J_i(z)}{dz} = 0, \quad 0<z<L,  \quad  i=1,2, \ldots, n,
\end{equation}
with $\J_i$ denoting the flux density function of ionic species $i$ as defined by
\begin{equation}
\label{Ji-1D}
  \J_i(z) =  - {\cal D}_i \left[Z_i c_i(z) \frac{du(z)}{dz} + \frac{dc_i(z)}{dz}  \right].
\end{equation}
In this work, $ \J_i$ is a scale function because both  $u$ and $c_i$ are the functions of one variable $z$.

We now  use the above equations  to derive the GHK equations by assuming that
 \[ \rho(z) = 0, \qquad  \sum_{i=1}^{n}Z_{i} c_i(z) = 0, \quad  0<z<L.\]
In other words, none of charges are considered within an ion channel pore interval. Under this assumption, the Poisson equation \eqref{electricField-Def} becomes independent of the steady Nernst-Planck equation \eqref{NP1D0}. Thus, a two-point boundary value problem of $u$ is produced from model \eqref{1DPNPionChannel}  as follows:
\begin{equation*}
  \left\{ 
\begin{array}{ll}
 \frac{d^2 u(z)}{dz^2} =0, & 0<z<L\\
 u(0)=u_0, \quad  u(L)=u_L. &  
\end{array}
\right.
\end{equation*}
The solution $u$ of the above boundary value problem can be easily found in the expression 
\begin{equation}
\label{u-def}
  u(z) = \frac{u_L - u_0}{L} z + u_0, \quad 0\leq z \leq L,
\end{equation}
resulting in a constant electric field in the form 
\begin{equation}
\label{A-def}
   \frac{d u(z)}{dz}  = - \frac{V_m}{L},
\end{equation}
where $V_m$ is defined by 
\begin{equation}
\label{V-1D}
  V_m=u_0 - u_L,
\end{equation} 
which will be referred to as a membrane potential (or a voltage across a cell membrane). 

From \eqref{NP1D2} it implies that the flux function $ \J_i(z)$ is either 0 or a nonzero constant. In 
 the case in which $\J_i=0$, we use \eqref{Ji-1D} to get 
\[   Z_i c_i(z) \frac{du(z)}{dz} + \frac{dc_i(z)}{dz}= 0,\quad 0<z<L. \]
By separating the functions $u$ and $c_i$, the above equation is rewritten as
\[ \frac{1}{c_i} \frac{d c_i}{dz} = - Z_i  \frac{du}{dz}.\]
Integrating over the interval $(0, L)$ on the both sides of the above equation, we obtain the Nernst equation:
\begin{equation}
\label{NP-def}
     V_m = \frac{1}{Z_i} \ln \frac{c_i(L)}{c_i(0)}, \quad  i=1,2, \ldots, n.
\end{equation}
 
We next consider the case in which $\J_i$ is a constant. Since the electric field $ \frac{du}{dz}$ is a constant, as given in  \eqref{A-def}, we can use \eqref{Ji-1D} to construct a two-point boundary value problem for determining the concentration $c_i$ of ionic species $i$ as follows:
\begin{equation}
\label{JiCi1D}
\displaystyle \left\{
\begin{array}{ll}
\displaystyle \frac{d c_i(z)}{dz} - Z_i  \frac{V_m}{L} c_i(z)  = - \frac{\J_i}{{\cal D}_i}, &0<z<L, \\
c_i(0) =c_{i,0}, \quad  c_i(L) =c_{i,L}. &
\end{array}
\right.
\end{equation}
Since $ \frac{\J_i}{{\cal D}_i}$ is a constant,  by the integrating factor method, $c_i$ can be found in the expression
\begin{equation}
\label{ci-formula0}
    c_i(z) = \frac{1}{\theta(z)} \left[  - \frac{\J_i}{{\cal D}_i} \int_{Z1}^z \theta(z) dz + c_{i,0} \right], 
\end{equation}
where $\theta(z)$ is an exponential integrating factor, which can be found as follows:
\[     \theta(z) = e^{- \int_0^z Z_i \frac{V_m}{L}  dz} = e^{-Z_i \frac{V_m}{L} z}, \]
where $V_m$ has been defined in \eqref{V-1D}. 
Applying the above expression to \eqref{ci-formula0}, we get
\begin{equation}
\label{ci-formula1}
    c_i(z) = e^{Z_i \frac{V_m}{L} z} \left[ \frac{\J_i L}{{\cal D}_i Z_iV_m} \left( e^{-Z_i \frac{V_m}{L} z} -1 \right)+ c_{i,0} \right].
\end{equation}
We then set $z=L$  on the both sides of \eqref{ci-formula1} to get a linear equation of $\J_i$: 
\[     c_i(L) =   e^{Z_i V_m }\left[ \frac{\J_i L}{{\cal D}_i Z_iV_m} \left( e^{-Z_i V_m } -1 \right)+ c_{i,0} \right]. \]
Solving the above equation for $\J_i$, we obtain an expression of $\J_i$ as follows:
\begin{equation}
\label{GHK-eq0}
    \J_i = Z_i {\cal D}_i \left(\frac{V_m}{L} \right) \frac{ c_{i,0} - c_{i,L} e^{-Z_i V_m}  }{1 - e^{-Z_i V_m}}, \qquad  i=1,2,\ldots, n,
\end{equation}
which can also be rewritten as
\begin{equation}
\label{GHK-eq1}
    \J_i = Z_i {\cal D}_i \left(\frac{V_m}{L} \right)  \frac{c_{i,L}  -  c_{i,0} e^{Z_i V_m}  }{1 - e^{Z_i V_m} }.
\end{equation}
The expression \eqref{GHK-eq0} or \eqref{GHK-eq1} is referred to as {\bf the GHK flux equation} in the literature.

The electrical  current ${\cal I}_{i}$ of species $i$ can be derived from the multiplication of  $\J_i$ by the charge $Z_ie_c$ of species $i$:
\[ {\cal I}_{i} = Z_i e_c  \J_i, \quad i=1,2,\ldots, n.\]
Using the GHK flux equation \eqref{GHK-eq0}, we immediately derive {\bf the GHK electric current equation}:
\begin{equation}  
 \label{currents}
   {\cal I}_{i} = Z_i^2 e_c {\cal D}_i \left(\frac{V_m}{L} \right) \frac{ c_{i,0} - c_{i,L} e^{-Z_i V_m}  }{1 - e^{-Z_i V_m}}.
\end{equation}  
Adding the above currents together, we obtain  a net electrical current, denoted by ${\cal I}$,  as follows:
\begin{equation}  
 \label{net_current}
   {\cal I} =   \sum_{i=1}^n  {\cal I}_{i} = e_c \frac{V_m}{L} \sum_{i=1}^n Z_i^2  {\cal D}_i \frac{ c_{i,0} - c_{i,L} e^{-Z_i V_m}  }{1 - e^{-Z_i V_m}}.
\end{equation}  
 If the net current is zero, setting $ {\cal I}=0$, we derive   {\bf the GHK voltage equation}:
 \begin{equation}  
 \label{GHK_potential}
    \sum_{i=1}^n Z_i^2  {\cal D}_i \frac{ c_{i,0} - c_{i,L} e^{-Z_i V_m}  }{1 - e^{-Z_i V_m}} = 0.
 \end{equation}   
 A solution $V_m$ of the above equation  is called the GHK potential.
 
For example, for a solution with sodium (Na$^+$ with $Z_1=1$), potassium (K$^+$ with $Z_2=1$), and chloride (Cl$^-$ with $Z_3=-1$) ions ($n=3$) \cite[Eq. (2.3)]{fall2002computational} and \cite[Eq. (2.7.2)]{keener1998mathematical}, we can solve the equation  \eqref{GHK_potential} for $V_m$ to get the GHK potential in the expression:
\[      V_m = -\ln \frac{ {\cal D}_1 c_{1,0} +  {\cal D}_2 c_{2,0}+  {\cal D}_3 c_{3,L}}
                                 { {\cal D}_1 c_{1,L} +  {\cal D}_2 c_{2,L} +  {\cal D}_3 c_{3,0}}, \]
 where $c_{1,0}, c_{2,0}$ and $c_{3,0}$ are the concentration values of sodium, potassium, and chloride ions at $z=0$ while
 $c_{1,L}, c_{2,L}$ and $c_{3,L}$  at $z=L$, respectively.

A  membrane potential, $V$,  in volts is defined by
\[     V= \Phi(0) - \Phi(L).\]
Using \eqref{u_Phi} and \eqref{V-1D}, we can derive $V$ from $V_m$ by
\begin{equation}
\label{V_Vm}
    V = \frac{k_B T}{e_c} V_m.
\end{equation}

Using  \eqref{GHK-eq0} and \eqref{V_Vm}, we can obtain another expression of the GHK  flux equation as follows: 
\begin{equation}
\label{GHK-eq}
    \J_i = Z_i {\cal D}_i \left( \frac{e_c }{k_B T L}\right) V \frac{c_{i,0}  - c_{i,L}e^{-\frac{Z_ie_c}{k_B T} V} }{1 -  e^{-\frac{Z_ie_c}{k_B T} V}}, \quad i=1,2, \ldots, n.
\end{equation}
As a special case, when $c_{i,0}=c_{i,L} = c_i^b$, the above expression is simplified as
\[    \J_i = Z_i {\cal D}_i c_i^b  \left( \frac{e_c }{k_B T L}\right) V,\]
indicating that the flux $\J_i$ has a linear relationship with the membrane potential $V$. In other words, the GHK  flux equation describes a nonlinear relationship of $\J_i$ with $V$ only if  $c_{i,0} \neq c_{i,L}$.

\section{An extension of GHK equations}
In this section, we extend the GHK equations from a constant electric field to a nonlinear electric field based on the 1D PNPic model \eqref{1DPNPionChannel} with a nonzero permanent charge function $\rho$. Because of \eqref{NP1D2}, we still have that the flux $\J_i$ of species $i$ is either zero or a nonzero constant. Since the case of $\J_i$ being zero can be treated in the same way  as what is done in the previous section, we only consider the case of $\J_i$ being a nonzero constant in this section.

Similarly to what is done in the derivation of GHK flux equation, we treat the nonlinear electric field $du/dz$ as a known coefficient function  of the steady Nernst-Planck equation \eqref{NP1D0}. In this way, we can separate  \eqref{NP1D0} from the Poisson equation \eqref{electricField-Def} and then generate a two-point boundary value problem for each ionic concentration function, $c_i$, from  the 1D PNPic model \eqref{1DPNPionChannel} as follows:
\begin{equation*}
\label{JiCi1D2}
\displaystyle \left\{
\begin{array}{ll}
\displaystyle  \frac{d c_i}{dz} + Z_i  \frac{d u}{dz} c_i(z)  = - \frac{\J_i}{{\cal D}_i}, &0<z<L, \\
c_i(0) =c_{i,0}, \quad  c_i(L) =c_{i,L}. &
\end{array}
\right.
\end{equation*}
By the integrating factor method, the solution $c_i$ of the above problem can be found in the form
\begin{equation}
\label{new-ci-formula0}
    c_i(z) = \frac{1}{\theta(z)} \left[  - \frac{\J_i}{{\cal D}_i} \int_{0}^z \theta(\xi) d\xi + c_{i,0} \right],   \quad 0<z<L,
\end{equation}
where $\theta(z)$ denotes an exponential integrating factor, which can be found as shown below:  
\[  \displaystyle   \theta(z) = e^{\int_0^z Z_i \frac{du(\xi)}{d\xi}  d\xi} = e^{Z_i [u(z)- u_0]}. \]
Applying the above expression    to \eqref{new-ci-formula0}, we obtain an integral expression of  $c_i$ as follows:
\begin{equation*}
\label{new-ci-formula1}
    c_i(z) = e^{-Z_i [u(z)- u_0]} \left[  - \frac{\J_i}{{\cal D}_i} \int_0^z e^{Z_i [u(\xi)- u_0]} d\xi + c_{i,0} \right].
\end{equation*}
Setting $z=L$  on the both sides of the above expression, we get a linear equation of $\J_i$: 
\[     c_{i,L}=   e^{-Z_i [u_L- u_0]} \left[  - \frac{\J_i}{{\cal D}_i}  \int_0^L e^{Z_i [u(\xi)- u_0]} d\xi + c_{i,0} \right].\]
Solving the above equation for $\J_i$, we get  {\bf an extension of the GHK flux equation} in the form
\begin{equation}
\label{Ji-formula0}
    \J_i = {\cal D}_i \frac{ c_{i,0} -  c_{i,L}  e^{Z_i [u_L- u_0] }} {\int_0^L e^{Z_i [u(z)- u_0]} dz }, \quad i=1,2,\ldots, n,
\end{equation}

Clearly, the integral $\int_0^L e^{Z_i [u(z)- u_0]} dz$ can be reformulated as
\[   \int_0^L e^{Z_i [u(z)- u_0]} dz = e^{-Z_i u_0} \int_0^L e^{Z_i u(z)} dz.\]
Note that the integral $ \int_0^L e^{Z_i u(z)} dz$  is a constant that collects all the values of potential function $u$ over the ion channel interval $[0,L]$. Hence, we can use it to define an extension parameter, $\alpha_i$,  by
\begin{equation}
\label{alpha-def}
   \alpha_i = \int_0^L e^{Z_i u(z)} dz, \quad i=1,2,\ldots,n.
\end{equation}

For clarity, we denote the extended GHK flux $\J_i$  of \eqref{Ji-formula0} by the new notation $\J_{i}^E$ and use  the extension parameter $\alpha_i$ to reformulate it as follows:
\begin{equation}
\label{Ji-formula1}
    \J_i^E ={\cal D}_i \frac{e^{Z_iu_0}}{\alpha_i} \left[ c_{i,0} - c_{i,L}  e^{-Z_i V_m }\right], \quad i=1,2, \ldots, n,
\end{equation}
where $V_m$ and $\alpha_i$ have been defined in \eqref{V-1D} and \eqref{alpha-def}, respectively.

Following what is done in \eqref{currents}, we can obtain {\bf an extened GHK current equation} as follows:
\begin{equation}  
 \label{extension_currents}
     {\cal I}_{i}^E =  e_c Z_i {\cal D}_i \frac{e^{Z_iu_0}}{\alpha_i}  \left[ c_{i,0} -  c_{i,L} e^{-Z_i V_m } \right], \quad i=1,2,\ldots, n, 
 \end{equation} 
 where ${\cal I}_{i}^E$ denotes an electrical  current of species $i$. 
  
 Adding all ${\cal I}_{i}^E$ together, we obtain a  net electrical current, ${\cal I}^E$, in the expression:
\begin{equation}  
 \label{extension_net_current}
     \quad {\cal I}^E = e_c  \sum_{i=1}^n Z_i  {\cal D}_i \frac{e^{Z_iu_0}}{\alpha_i}  \left[ c_{i,0} -  c_{i,L}  e^{-Z_i V_m } \right].
 \end{equation} 
 
 Similar to what is done in \eqref{GHK_potential}, we can set ${\cal I}^E =0$ to obtain  {\bf  an extended GHK voltage equation} as follows: 
 \begin{equation}  
 \label{extension_GHK_potential}
      \sum_{i=1}^n Z_i {\cal D}_i \frac{e^{Z_iu_0}}{\alpha_i} \left[ c_{i,0} - c_{i,L} e^{-Z_i V_m } \right]= 0.
 \end{equation} 
 A solution of the above equation gives {\bf an extended ned GHK potential}. 
 
 As an example, from the equation of  \eqref{extension_GHK_potential} we can get an extended ned GHK potential, denoted by $V_m^{E}$,  for an ionic solution with sodium (Na$^+$ with $Z_1=1$) and chloride (Cl$^-$ with $Z_2=-1$) ions ($n=2$) in the expression
\begin{equation}  
 \label{extension_GHK_potential_ex1}
     V_m^{E} = - \ln \frac{ \varpi + \sqrt{\varpi^2 + 4 \frac{ \alpha_1}{\alpha_2} {\cal D}_1{\cal D}_2 e^{-2u_0} c_{1,L} c_{2,L}}}{2 {\cal D}_1 c_{1,L} },
 \end{equation}
 where $c_{1,L}$ and $c_{2,L}$ denote the concentration values of sodium and chloride ions at $z=L$,  respectively; ${\cal D}_1$ and ${\cal D}_2$ are the diffusion constants of sodium and chloride ions,  respectively; $\alpha_1$, $\alpha_2$, and $\varpi$ are defined by
 \[ \alpha_1 = \int_0^L e^{u(z)} dz, \quad \alpha_2 = \int_0^L e^{-u(z)} dz, \]
 and
 \[     \varpi = {\cal D}_1 c_{1,0} -  \frac{ \alpha_1}{\alpha_2} {\cal D}_2 e^{-2u_0} c_{2,0}.\]
Here $\varpi$ has been assumed to be nonnegative and we have selected the positive square root  in \eqref{extension_GHK_potential_ex1} to ensure that the extended ned GHK potential $V_m^{E}$ is well-defined.

\section{Numerical calculation for extension parameter}
In this section, we present a numerical quadrature scheme for computing the extension parameter $\alpha_i$ approximately. In this scheme, we assume that a set of interpolation points, ${\cal S}_m$, is  given by
\begin{equation}
\label{data_set}
 {\cal S}_m = \{ (z_j, u_j) \; | \; j=0, 1,2, \ldots, m\}, \quad m\geq 1,
\end{equation}
where  $z_j$ denotes the $j$-th partition number of  the ion channel pore interval $[0, L]$ satisfying 
\begin{equation}
\label{mesh_partition}
  0 =  z_0 < z_1 < z_2 < \ldots < z_{m-1} < z_m = L,
\end{equation}
and $u_j$ denotes a numerical value of an electrostatic potential function, $u$, at $z=z_j$ for $ j= 0, 1, 2, \ldots, m$. To simplify the presentation of our numerical scheme, we set $z_j=jh$ with the mesh size $h=L/m$.

When $u$ is nonlinear, we use the interpolation point set ${\cal S}_m$ to construct a piecewise linear interpolating function, $\hat{u}$, of $u$  in the expression
\begin{equation}
\label{piecewise_linear}
    \hat{u}(z) = a_j z + b_j \quad \mbox{for }  z_{j-1} \leq z \leq z_j,
\end{equation}
where $a_j = (u_j - u_{j-1})/h$ and $ b_j = (z_j u_{j-1} - z_{j-1} u_j) / h$ for $ j= 1, 2, \ldots, m$.
That is, $\hat{u}$ is a linear function within sub-interval $[ z_{j-1}, z_j]$.
We then estimate the extension parameter $\alpha_i$ approximately as shown below:
\begin{eqnarray*}
 \alpha_i &=& \int_0^L e^{Z_i u(z)} dz = \sum_{j=1}^m \int_{z_{j-1}}^{z_j} e^{Z_i u(z)} dz   \nonumber \\
                                    & \approx &  \sum_{j=1}^m \int_{z_{j-1}}^{z_j} e^{Z_i \hat{u}(z)} dz  
                                    =  \sum_{j=1}^m  \int_{z_{j-1}}^{z_j} e^{Z_i (a_j z + b_j ) } dz \nonumber \\
                             & = & \frac{1}{Z_i}  \sum_{j=1}^m \frac{e^{Z_i b_j }}{a_j} \left[ e^{Z_i a_j z_j } - e^{Z_i a_j z_{j-1} } \right]\\
                             & = & \frac{h}{Z_i} \sum_{j=1}^m \frac{e^{Z_i u_j} - e^{Z_i u_{j-1}} }{ u_j - u_{j-1}},
\end{eqnarray*}
where we have used the partition formula $z_j = z_{j-1} + h$.
For clarity, the above approximate value of $\alpha_i$  is denoted by the notation $\hat{\alpha}_i$. That is,
\begin{equation}
\label{integral_scheme}
   \hat{\alpha}_i  = \frac{h}{Z_i} \sum_{j=1}^m \frac{e^{Z_i u_j} - e^{Z_i u_{j-1}} }{ u_j - u_{j-1}}.
\end{equation}
This gives the formula for computing the extended GHK flux $\J_{i}^E$ approximately as follows:  
 \begin{equation}
\label{extend_GHK-eq}
   \J_i^E \approx {\cal D}_i \frac{e^{Z_iu_0}}{ \hat{\alpha}_i } \left[ c_i(0) -  c_i(L)  e^{-Z_i V_m }\right], \quad i=1,2,\ldots,n,
\end{equation}
where $V_m$ has been defined in \eqref{V-1D}.
Specifically, when $m=1$, we have  
\[ h=L, \quad z_0=0, \quad z_1=L, \quad  u_0= u(0), \quad u_1=u(L),\]
and
\[ \hat{\alpha}_1=\frac{h}{Z_i} \frac{e^{Z_i u_1} - e^{Z_i u_0} }{ u_1- u_0},\]
from  which \eqref{extend_GHK-eq}  can be reduced to  the GHK flux equation \eqref{GHK-eq0}. In this sense, our extended GHK flux equation contains the classic GHK flux equation as a special case (i.e. only using the two boundary values $u_0$ and $u_L$ of $u$). 

\vspace{6mm}

\section{A finite element iterative scheme for solving\\ 1DPNPic model}
In order to generate an interpolation date set of \eqref{data_set}, we present a nonlinear finite element iterative scheme for solving the 1DPNPic model \eqref{1DPNPionChannel}  in this section.  

We start with a finite element approximation of the 1DPNPic model \eqref{1DPNPionChannel}.

Let $V$ denote a linear finite element function space based on the mesh partition \eqref{mesh_partition}  of the interval $[0, L]$. Here each function of $V$ is linear within each subinterval of the partition and continuous on the interval $[0, L]$.  We define a subspace, $V_0$, of $V$ by
\[     V_0=\{v\in V \; | \; v(0) =0, \quad v(L)=0\}. \]
In the finite element method, $V$ and $V_0$ are referred to as the trial and test function spaces, respectively. 
Multiplying each equation of the system \eqref{1DPNPionChannel} by a test function, $v$, and integrating the second-order terms by parts, we can approximate  the 1DPNPic model \eqref{1DPNPionChannel}  as a system of nonlinear finite element variational problems:
Find $u \in V$ and $c_i \in V$ satisfying the Dirichlet boundary value conditions $u(0)=u_0$, $u(L)=u_L$, $c_i(0)=c_{i,0}$, and $c_i(L)=c_{i,L}$ such that
\begin{subequations}
\label{fem_equations}
\begin{eqnarray}
\label{fem4u}
  \es  \int_0^L \frac{du}{dz} \frac{dv}{dz} dz   -  \frac{e_{c}^{2}}{\ez k_{B}T}  \int_0^L  \left( \sum_{j=1}^{n}Z_{j}  c_j   \right) v  dz =  \int_0^L \rho(z) v  dz \quad \forall v \in V_0,\\
  \label{fem4c}
    \int_0^L {\cal D}_i \left( Z_i c_i  \frac{du}{dz}    + \frac{dc_i}{dz} \right) \frac{d v_i}{dz}  dz  =0  \quad \forall v_i \in V_0, \quad  i=1,2,\ldots,n.
\end{eqnarray}
\end{subequations}

We next present a damped iterative scheme for solving the above nonlinear finite element system.

Let $u^{(k)}$ and  $c_i^{(k)}$ denote the $k$th iterates of $u$ and $c_i$, respectively. When the initial iterates $u^{(0)}$  and $c_i^{(0)}$ are given, for $k= 0, 1,2,\ldots,$ we define the updates $u^{(k+1)}$ and  $c_i^{(k+1)}$ by
\begin{subequations}
\label{iteration_scheme}
\begin{eqnarray}
\label{cj-iterate}
   c_i^{(k+1)} &=&  c_i^{(k)} + \omega( {p}_i  - c_i^{(k)}),  \quad i=1, 2, \ldots,n, \\
    \label{u-iterate}
     u^{(k+1)} &=& u^{(k)} +  \omega ({q} - u^{(k)} ),
\end{eqnarray}
\end{subequations}
where $\omega$ is a damping parameter between 0 and 1, ${p}_i$ is a solution of the linear finite element variational problem: Find $p_i \in V$ satisfying $p_i(0)=c_{i,0}$ and $p_i(L)=c_{i,L}$ for $ i=1, 2,  \ldots,n$ such that
\begin{equation}
    \label{pi-def}
     \int_0^L {\cal D}_i \left( Z_i p_i  \frac{d u^{(k)}}{dz}   +  \frac{d p_i}{dz}  \right) \frac{d v_i}{dz}  dz  =0  \quad \forall v_i \in V_0, 
\end{equation} 
and ${q}$ is a solution of  the linear finite element variational problem: Find  $q \in V$ satisfying $q(0)=u_0$ and $q(L)=u_L$ such that
\begin{equation}
    \label{q-def}
\es   \int_0^L  \frac{d q}{dz}  \frac{d v}{dz}  dz   =  \frac{e_{c}^{2}}{\ez k_{B}T}  \int_0^L \left( \sum_{j=1}^{n}Z_{j}  c_j^{(k+1)}   \right) v  dz     
                                                                          +  \int_0^L \rho(z) v  dz \quad \forall v \in V_0.
\end{equation} 
Note that we have separate the nonlinear system \eqref{fem_equations} into $n+1$ independent linear equations by substituting the current iterate $u^{(k)}$ to $u$ in \eqref{fem4u} and the updates $c_i^{(k+1)}$ to $c_i$ in \eqref{fem4c}.
By default, the initial iterates $u^{(0)}$  and $c_i^{(0)}$ are set by
\begin{equation}
    \label{initial_iterates}
   u^{(0)} = \frac{u_L - u_0}{L} z + u_0, \quad c_i^{(0)}=0, \quad  i=1,2,\ldots, n, 
\end{equation} 
 since the above selection can lead to the GHK flux equation as shown in Section~2.
We terminate the iterative process whenever the following iteration termination rule holds:
\begin{equation}
    \label{Ite-stop}
    \|   u^{(k+1)}  -  u^{(k)}  \| < \epsilon,  \quad  \max_{1\leq i\leq n} \|   {c}^{(k+1)}_i -  {c}^{(k)}_i  \| < \epsilon,
\end{equation} 
where $\epsilon$ is a tolerance (e.g. $\epsilon=10^{-5}$) and  $ \| \cdot \|$ is defined by  
\[     \|v\| = \sqrt{\int_0^L |v(z)|^2 dz} \quad \mbox{for }v\in V. \]

\begin{table}
\begin{centering}
\scalebox{1.0}{
    \begin{tabular}{|c|c|c|c|c|c|c|c|}
      \hline
       & $\omega$       & Iter.   & $\J_1^E$  & $\J_2^E$     & $\frac{|\J_1^E - \J_{1}| }{|\J_1^E |}$ & $\frac{|\J_2^E - \J_{2}| }{|\J_2^E|}$    & CPU time  (sec.)     \\  \hline
 Test 1      & 0.2     &    59 & $ -0.02590$ & 0.00039 &  0.7394 &  3.7888 & 0.3237  \\ \hline                                                                        
Test 2        & 0.2    & 58  &  $-0.00082$  &  0.01196  &   7.2532 &  0.8430   &  0.3142   \\ \hline  
Test 3        &  0.19   & 57  & $-0.00084$ & 0.00664   &  7.0130 &  0.7172  & 0.2712 \\ \hline   
Test 4      & 0.6     &    14 & $ -0.00499$ & 0.00255  &  0.3539 &  0.2628 & 0.1385  \\ \hline                                                                        
Test 5      & 0.6    & 11  &  $-0.00133 $  &  0.00203   &   $2.61\times 10^{-15}$ & $3.63\times 10^{-15}$   &0.0881   \\ \hline                                         
 \end{tabular}}
 \caption{A comparison of the fluxes  $\J_1^E$ and $\J_2^E$  calculated by our extended GHK equations \eqref{extend_GHK-eq} with the fluxes  $\J_1$ and $\J_2$ calculated by the classic GHK equations \eqref{GHK-eq}, along with the convergence and performance of our finite element iterative scheme \eqref{iteration_scheme}. Here $\J_1 \approx -0.00675$ and $\J_2 \approx  0.00188$ for the five tests.}
     \label{performance}
 \end{centering}
\end{table} 

\begin{figure}
 \centering
       \begin{subfigure}[b]{0.28\textwidth}
                \centering
                \includegraphics[width=\textwidth]{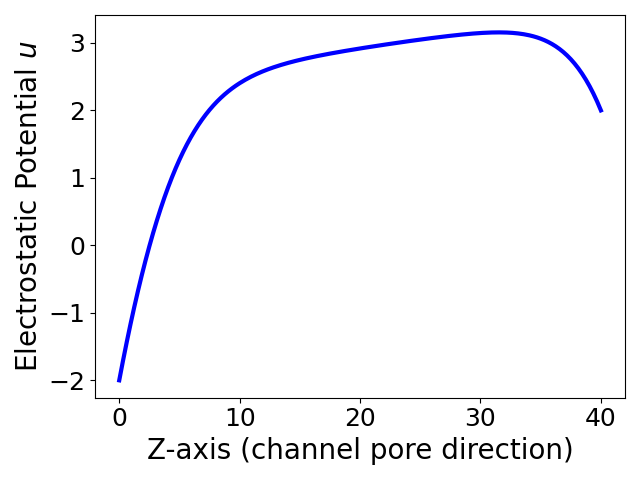}
        \end{subfigure} 
        \qquad
         \begin{subfigure}[b]{0.28\textwidth}
                \centering
                \includegraphics[width= \textwidth]{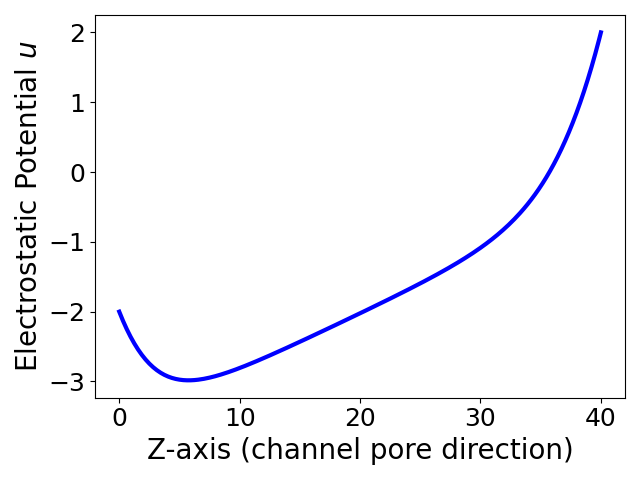}
        \end{subfigure}     
        \qquad
         \begin{subfigure}[b]{0.28\textwidth}
                \centering
                \includegraphics[width= \textwidth]{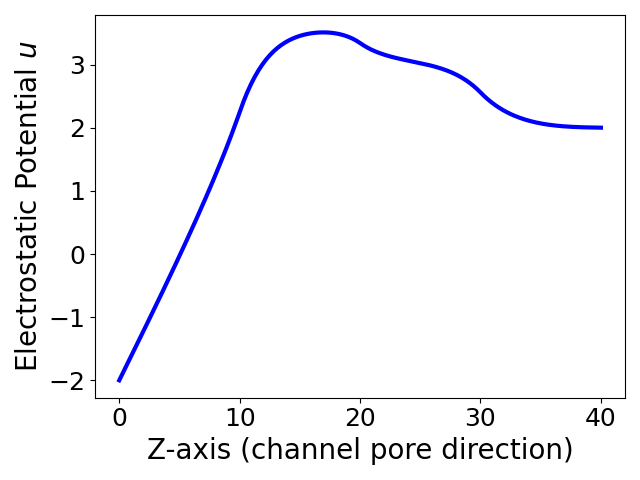}
        \end{subfigure} 
        \begin{subfigure}[b]{0.28\textwidth}
                \centering
                \includegraphics[width= \textwidth]{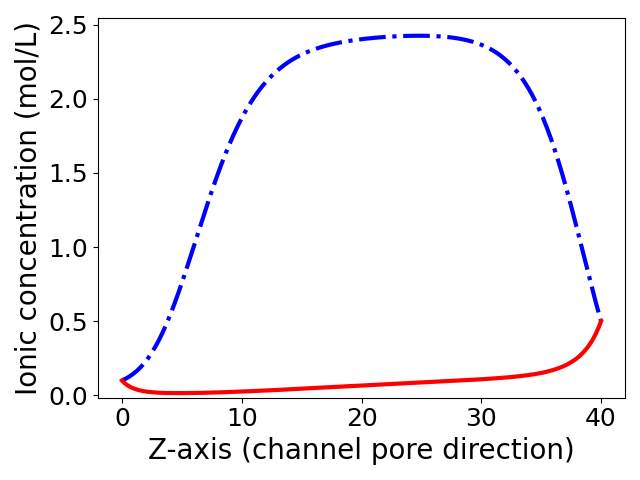}
                \caption{Test 1 case}  
        \end{subfigure}   
        \qquad
        \begin{subfigure}[b]{0.28\textwidth}
                \centering
                \includegraphics[width=\textwidth]{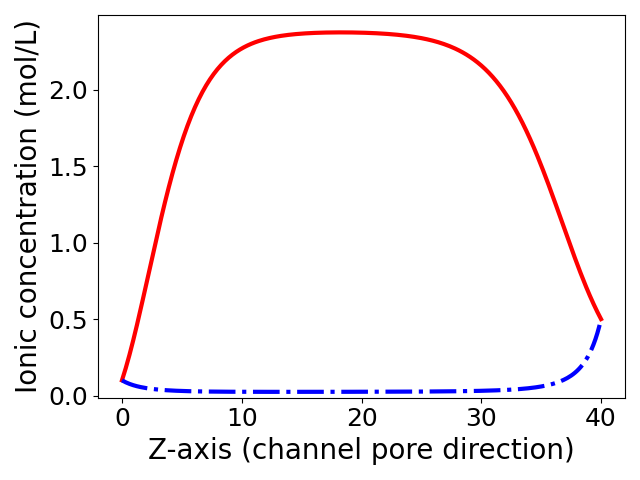}
                 \caption{Test 2 case}  
        \end{subfigure}   
       \qquad 
        \begin{subfigure}[b]{0.28\textwidth}
                \centering
                \includegraphics[width= \textwidth]{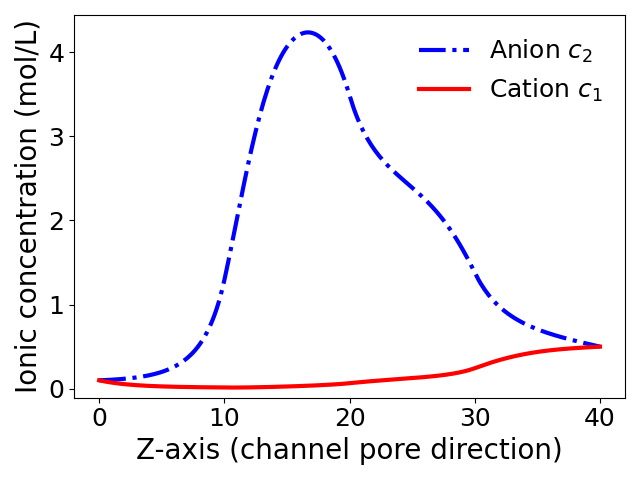}
                 \caption{Test 3 case}  
        \end{subfigure}        
                \caption{The electrostatic potential function $u$ and ionic concentrations $c_1$ and $c_2$  generated from Tests 1, 2, and 3 by our finite element program package for solving the 1DPNPic model \eqref{1DPNPionChannel}. }         
                \label{Compare_concentrations}
\end{figure}

\section{Numerical results}
We implemented the damped iterative scheme \eqref{iteration_scheme} in Python as a program package based on the state-of-the-art finite element library from the FEniCS project \cite{fenics-book}. In numerical tests, we used a piecewise constant  permanent charge function,  $ \rho$, in the expression
\begin{equation}
 \rho(z) = \left\{
\begin{array}{ll}
 Q_1, &  0  < z \leq 10,\\
 Q_2, &  10  < z \leq 20,\\
 Q_3, &  20  < z \leq 30,\\
 Q_4, &  30  < z \leq 40,\\
 \end{array}
\right.
\end{equation}
where $Q_i$ is a net charge within the $i$th sub-interval of the ion channel interval $[0, L]$ with $L=40$ for $i=1,2,3,4$. When all $Q_i$ are equal to $Q$, $\rho$ becomes a constant function. We did three numerical tests, called Tests 1, 2, and 3,  using the following values of $Q$ and $Q_i$:
\begin{description}
\item[Test 1] Set  $Q_i= Q$ for $i=1,2,3,4$ with $Q=10$.
\item[Test 2] Set $Q_i= Q$  for $i=1,2,3,4$ with  $Q=-20$.
\item[Test 3] Set $Q_1 = 0,  Q_2 =20,  Q_3 = 10,$ and $Q_4 = 0$. 
\end{description}
Here Tests~1 and 2 are used to study the extended GHK flux equation and 1DPNPic model within an anion-selectivity filter and a cation-selectivity filter, respectively. In Test~3, the interval $[0, L]$ contains  a selectivity filter (i.e. $10< z \leq 30$) and two solution regions (i.e.  $(0, 10]$ and $(30, 40]$), and  $\rho$ is a  non-constant positive charge function within the filter and zero within the two neutral solution regions.  See Figure~\ref{channel_pore_intervals}(a) for an illustration of the ion channel interval $[0, L]$ used in Tests 1 and 2 and Figure~\ref{channel_pore_intervals}(b) for that used in Test 3. 

We also did numerical tests without considering any permanent charge effect as follows: 
\begin{description}
\item[Test 4] Set  $\rho=0$ and  $c_{i,0}\neq c_{i,L}$ with $c_{i,0}=0.1$ and $c_{i,L}=0.5$ for $i=1,2$.
\item[Test 5] Set $\rho=0$ and  $c_{i,0} = c_{i,L}$ with $c_{i,0}=0.1$ and $c_{i,L}=0.1$ for $i=1,2$.
\end{description}

\begin{figure}[t]
 \centering
       \begin{subfigure}[b]{0.35\textwidth}
                \centering
                \includegraphics[width=\textwidth]{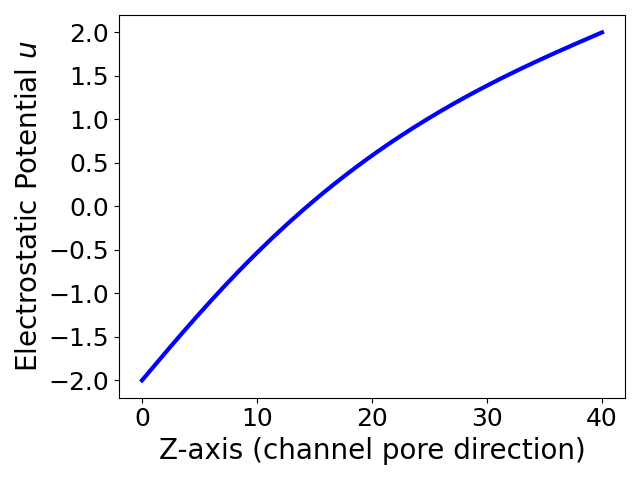}
        \end{subfigure}    
        \qquad  \qquad 
        \begin{subfigure}[b]{0.35\textwidth}
                \centering
                \includegraphics[width= \textwidth]{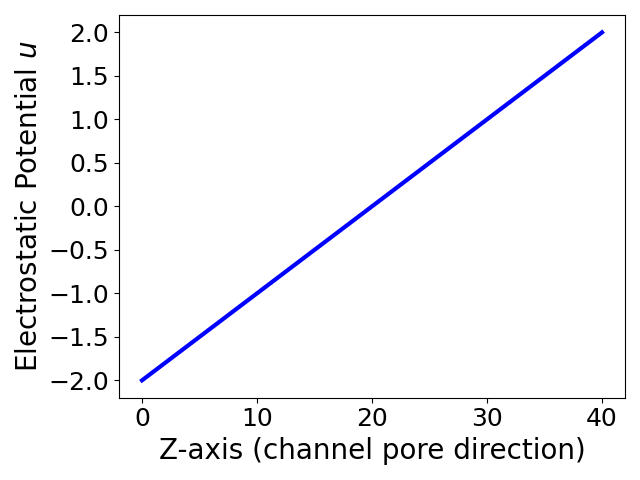}
        \end{subfigure}      
        \begin{subfigure}[b]{0.35\textwidth}
                \centering
                \includegraphics[width=\textwidth]{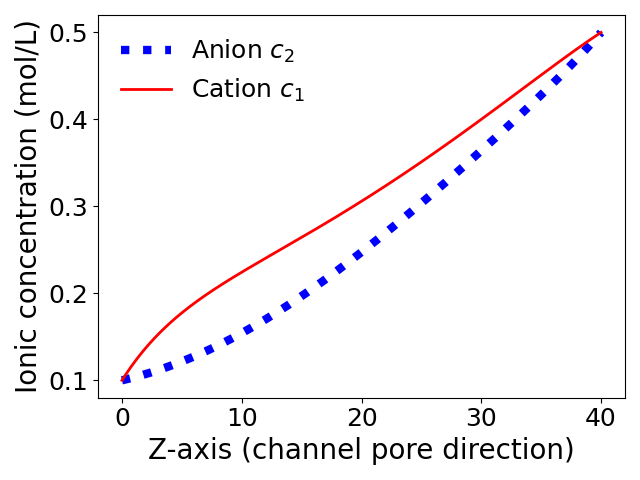}
                 \caption{Test 4 case}  
        \end{subfigure}   
        \qquad  \qquad 
        \begin{subfigure}[b]{0.35\textwidth}
                \centering
                \includegraphics[width= \textwidth]{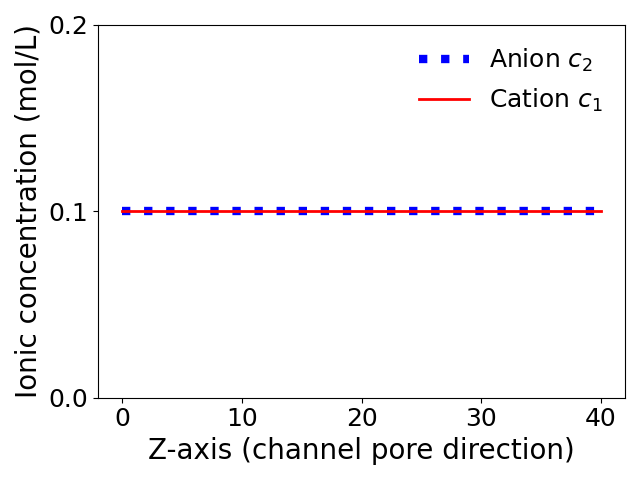}
                 \caption{Test 5 case}  
        \end{subfigure}   
                 \caption{The electrostatic potential functions of $u$ and ionic concentrations of $c_1$ and $c_2$  generated from Tests 4 and 5 by our finite element program package for solving the 1DPNPic model \eqref{1DPNPionChannel}. }         
                \label{results4tests4_5}
\end{figure}

In these numerical tests, the units of length and concentration were set as angstroms and moles per liter (mol/L), respectively; a mesh partition of \eqref{mesh_partition} was set with $m=256$ and $L=40$, which gave the mesh size $h \approx 0.156$; a solution of two ionic species (i.e. $n=2$) was used with $c_1$ denoting a concentration of cations and $c_2$  a concentration of anions; the boundary values of \eqref{boundary_values} were set as $u_0 =-2$, $u_L = 2,$ $c_{i,0}=0.1$, and $c_{i,L}=0.5$ for $i=1,2$. We also fixed the parameters  $\es=80$, $Z_1=1$, ${\cal D}_1= 0.133 $ (for Na$^+$), $Z_2=-1$, and ${\cal D}_2= 0.203$ (for Cl$^-$). The initial iterates were set by \eqref{initial_iterates}. Each finite element equation of \eqref{pi-def} and \eqref{q-def} were solved by the Gaussian elimination method.  The numerical tests were done on our iMac computer with one 4.2 GHz Intel core i7 processor and 64 GB memory.   The numerical results are reported in Table~\ref{performance} and Figures~\ref{Compare_concentrations} and~\ref{results4tests4_5}.

Table~\ref{performance} reports the flux values $\J_1^E$ and $\J_2^E$  calculated by our extended GHK flux equations \eqref{extend_GHK-eq}  and their relative errors with the ionic fluxes $\J_1$ and $\J_2$ calculated by the GHK flux equation  \eqref{GHK-eq0}. The relative error values can be large up to about 7, showing the importance of considering charge effects.
Table~\ref{performance}  also reports the convergence and performance of our nonlinear iterative scheme \eqref{iteration_scheme} in terms of the number of iterations and computer CPU time. Here the damping parameter $\omega$ was selected optimally (by numerical experiments) in the sense that  it enabled the number of iterations satisfying the   iteration termination rule \eqref{Ite-stop} to become the smallest. The computer CPU time was counted from the starting of an implementation to the completion of an iteration process. From the table it can be seen that our finite element iterative scheme \eqref{iteration_scheme} has a fast convergence rate and runs efficiently.  All the tests only took only 0.33 seconds or less in CPU time. 

Figure~\ref{Compare_concentrations} displays  the electrostatic potential functions and ionic concentration functions generated by our finite element program package for Tests 1, 2, and 3. In the case of Test 1, from Figure~\ref{Compare_concentrations}(a) it can be seen that anions have dominated cations due to a positive net charge within the selectivity filter of an anion  channel protein (such as a chloride channel). On the other hand, when a negative net charge is set in the selectivity filter of an ion channel protein, cations become to dominate anions as shown in Figure~\ref{Compare_concentrations}(b). This validates the selectivity of a cation channel (such as a potassium channel). In Test~3, we used a nonnegative piecewise constant function to mimic a charge distribution over a portion of an ion channel pore that contains one  anion selectivity filter and two neutral solution regions, resulting in a dominating concentration of anions as shown in Figure~\ref{Compare_concentrations}(c). These test results indicate that our finite element program package is a valuable tool in the study of the ion selectivity property of an ion channel protein.

Interestingly, the numerical results for Tests 4 and 5 reported in Table~\ref{performance} and Figure~\ref{results4tests4_5} show that the boundary values of ionic concentrations $c_i$ can significantly affect  the electrostatic potential and ionic distributions when $\rho=0$ (i.e. ignoring the charge effects from an ion channel protein). When the boundary values are different, as was done in Test~4, our GHK flux equation still produced different ionic fluxes from the classic GHK flux equation. However, with the same  boundary values, we got a linear potential function, $u$, of $z$ and two constant ionic concentrations $c_i$ for $i  =1,2$, as shown in Figure~\ref{results4tests4_5}(b). As a result, our  extended GHK flux equation produced the same ionic fluxes as the classic GHK equation.

\begin{table}
\begin{centering}
\scalebox{1.0}{
    \begin{tabular}{|c|c|c|c|c|c|}
      \hline
   $(m, h)$   & $\hat{\alpha}_1 $ & $\alpha_{1,CT}$   & $\alpha_1 $  & $\frac{| \hat{\alpha}_1  -  \alpha_1 | }{|\alpha_1  |}$ & $\frac{|\alpha_{1,CT} -\alpha_1  |}{| \alpha_1  |}$       \\  \hline
(32, 1.25)     &  581.434 & 583.129  &  581.433 & $1.76\times 10^{-6}$   & $2.92\times 10^{-3}$    \\ \hline      
(64, 0.625)     &  582.651 &  583.078  &  582.651   & $1.14\times 10^{-7}$   & $7.32\times 10^{-4}$    \\ \hline    
(128, 0.3125)     &  581.943  &  582.05   &   581.943 & $7.17\times 10^{-9}$   &  $1.83\times 10^{-4}$     \\ \hline    
(256, 0.15625)     &  581.24  &  581.267   &  581.24 & $4.49\times 10^{-10}$   & $4.58\times 10^{-5}$    \\ \hline         
\hline
   $(m, h)$   & $\hat{\alpha}_2 $ & $\alpha_{2,CT}$   & $\alpha_2 $  & $\frac{| \hat{\alpha}_2  -  \alpha_2 | }{|\alpha_2  |}$ & $\frac{|\alpha_{2,CT} -\alpha_2  |}{| \alpha_2  |}$       \\  \hline
(32, 1.25)     &  20.771 &  21.145   &   20.771  & $1.19\times 10^{-5}$   & $1.80\times 10^{-2}$    \\ \hline      
(64, 0.625)     & 20.616  & 20.711  &   20.616  & $7,77\times 10^{-7}$   & $4.59\times 10^{-3}$   \\ \hline    
(128, 0.3125)     &  20.544  &  20.568   &   20.544 & $4.95\times 10^{-8}$   & $1.16\times 10^{-3}$   \\ \hline    
(256, 0.15625)     &  20.509   &  20.515   &  20.509 & $3.12\times 10^{-9}$   & $2.90\times 10^{-4}$    \\ \hline                                          
 \end{tabular}}
 \caption{A comparison of our numerical quadrature value $\hat{\alpha}_i$ of \eqref{integral_scheme} with the composite trapezoidal value $\alpha_{i,CT}$ of \eqref{CT_formula} for Test 3. Here the extension parameter $\alpha_i$ of \eqref{alpha-def} is calculated by the FEniCS integral tool in high accuracy.
}
     \label{accuracies}
 \end{centering}
\end{table} 

Finally, we made numerical tests on our numerical quadrature \eqref{integral_scheme} to demonstrate its numerical accuracy improvement  in comparison to  the classical composite trapezoidal (CT) rule  in the expression \cite{NA_Burden_Faires}:
\begin{equation}
\label{CT_formula_0}
   \int_0^L g(z) dz \approx \frac{h}{2} \left[ g(0) + 2 \sum_{j=1}^{m-1} g(z_j) + g(L) \right]. 
\end{equation}
Using the partition \eqref{mesh_partition} of $[0, L]$ and $g =  e^{Z_i u}$, we can obtain the CT value $\alpha_{i,CT}$ of the extension parameter $\alpha_i$ for $ i=1,2, \ldots, n$ in the expression:
\begin{equation}
\label{CT_formula}
   \alpha_{i,CT} =  \frac{h}{2} \left[ e^{Z_i u(0)} + 2 \sum_{j=1}^{m-1} e^{Z_i u(z_j)} + e^{Z_i u(L)} \right].
\end{equation}
As the references for the accuracy comparison between our scheme and the CT rule, we also calculated the extension parameter $\alpha_i$ (i.e. the integral  $\int_0^L e^{Z_i u(z)} dz$) in high accuracy using the FEniCS integral computational tool \cite{fenics-book}.   We used Test 3 as the test problem since in this case,  the potential function $u$ is more nonlinear  than the other test cases (as shown in Figure~\ref{Compare_concentrations}). In the tests, we set $m=32, 64, 128, 256$ to get the mesh size $h=1.25, 0.625, 0.3125, 0.15625$, respectively. The test results are reported in Table~\ref{accuracies}. 

Table~\ref{accuracies} shows that our new numerical scheme for computing the extension parameter  has a higher degree of accuracy than the  composite trapezoidal rule. In fact, the composite trapezoidal rule \eqref{CT_formula} is constructed from a piecewise linear interpolation function of the composite function $e^{Z_i u}$  \cite{NA_Burden_Faires}. In contrast, our numerical quadrature \eqref{integral_scheme} is constructed from a piecewise linear interpolation function of $u$. Hence,  our new scheme can have much smaller relative errors than the classical composite Trapezoidal rule because its truncation error only comes from a piecewise linear interpolation function of $u$, instead of the composite function $e^{Z_i u}$, as was done in  the  composite Trapezoidal rule.

\section{Conclusions}
In this paper, we have extended the classic GHK equations from a constant electric field to a nonlinear electric field. The extended GHK equations have one additional parameter, called the extension parameter, to account for the charge effects from an ionic solution and an ion channel protein. For a constant electric field, a value of the extension parameter is found analytically such that our extended GHK equations are reduced to the classic GHK equations as special cases. When the electric field is nonlinear, we have developed a novel numerical quadrature scheme for computing the extension parameter approximately in terms of a set of electrostatic potential values. To this end, the extended GHK equations can be viewed as a bridge between the ``macroscopic" ion channel kinetics (i.e. ionic fluxes, electric currents, and membrane potentials) and  the ``microscopic" electrostatic potential values. Hence, the next step is to develop methodologies and computational tools for the generation of required electrostatic potential  values. 

One natural way to do so is to use a numerical solution of the 1D PNPic model since this model has been applied to the derivation of the extended GHK equations. In this paper, we developed a finite element iterative scheme for solving the 1D PNPic model and implemented it as a Python program package. Using this package, we generated different sets of electrostatic potential values and then did different numerical tests to study the extended GHK equations, the numerical quadrature scheme, and the 1D PNPic finite element solver. The numerical results from these numerical studies have confirmed the importance of considering charge effects in the calculation of ionic fluxes.  They also show that our numerical quadrature scheme has a higher degree of numerical accuracy than the classical composite trapezoidal rule. Furthermore, they  validate the convergence of our nonlinear finite element iterative scheme and demonstrate the high performance of our program package. This package  will be a valuable tool for the numerical study of an ion channel protein by itself. 

\section*{Acknowledgement}  
This work was partially supported by the Simons Foundation through research award number  711776 and the National Science Foundation, USA, through award number  DMS-2153376.


\end{document}